\documentclass[aps,
twocolumn,
superscriptaddress]{revtex4}%
\usepackage{amsfonts}
\usepackage{amsmath}
\usepackage{amssymb}
\usepackage{graphicx}%
\usepackage[dvipsnames]{xcolor}
\def\mJ{\mathcal J}

\begin{document}
%\preprint{HEP/123-qed}

\title{Magnetic exchange and susceptibilities in fcc iron: A supercell \\ dynamical
mean-field theory study}

\author{A. A. Katanin}
\affiliation{M. N. Mikheev Institute of Metal Physics, Russian Academy of Sciences, 620137 Yekaterinburg, Russia}
%\affiliation{Ural Federal University, 620002 Yekaterinburg, Russia}

\author{A. S. Belozerov}
\affiliation{M. N. Mikheev Institute of Metal Physics, Russian Academy of Sciences, 620137 Yekaterinburg, Russia}
\affiliation{Ural Federal University, 620002 Yekaterinburg, Russia}

\author{V. I. Anisimov}
\affiliation{M. N. Mikheev Institute of Metal Physics, Russian Academy of Sciences, 620137 Yekaterinburg, Russia}
\affiliation{Ural Federal University, 620002 Yekaterinburg, Russia}

\begin{abstract}
We study the momentum- and temperature dependencies of magnetic susceptibilities and magnetic exchange in paramagnetic fcc iron by a combination of density functional theory and supercell dynamical mean-field theory (DFT+DMFT). We find that in agreement with experimental results the antiferromagnetic correlations with the wave vector close to $(0,0,2\pi)$ dominate at low temperatures (as was also obtained previously theoretically), while the antiferromagnetic and ferromagnetic correlations closely compete at the temperatures $T\sim 1000$ K, where $\gamma$-iron exists in nature. Inverse staggered susceptibility has linear temperature dependence at low temperatures, with negative Weiss temperature $\theta_{\rm stagg} \approx -340$ K; the inverse local susceptibility is also linear at not too low temperatures, showing well formed local moments. Analysis of magnetic exchange shows that the dominant contribution comes from first two coordination spheres. In agreement with the analysis of the susceptibility, the nearest-neighbor exchange is found to be antiferromagnetic at low temperatures, while at temperature of the $\alpha$-$\gamma$ structural phase transition its absolute value becomes small, and the system appears on the boundary between the regimes with strongest antiferro- and ferromagnetic correlations. 

\end{abstract}
\maketitle

\section{Introduction}

Gamma- (face centered cubic) iron exists in nature in a relatively narrow temperature interval from 1185 to 1660 K. In this temperature interval it is known to show Curie-Weiss behavior of uniform magnetic susceptibility with large negative Weiss temperature \cite{Susc1,Susc2,Susc3}.
%short-range antiferromagnetic order
In Cu precipitates $\gamma$-iron
%\com{('The precipitates of $\gamma$-iron in copper'?) AK: I think both are possible, but the latter can not be combined with the remaining part of the sentence}
can be stabilized till very low temperatures, which allows studying its low-temperature magnetic properties. Early experimental studies have shown that this substance is a weak itinerant antiferromagnet with the Neel temperature of the order of $100$~K \cite{Neel1,Neel2,Neel3}. Later it was found \cite{Q1,Q2,Q3} that the corresponding incommensurate wave vector ${\bf Q} \approx
2\pi (1, 0.13, 0)$ in units of inverse lattice parameter $a$
is close to the so called AFM-I magnetic structure.  Therefore, in contrast to $\alpha$-iron, which possesses short-range ferromagnetic correlations above Curie temperature, $\gamma$-iron is expected to have short-range antiferromagnetic order above Neel temperature. 

The  stability of various ground states in $\gamma$-iron was analyzed theoretically within the density functional theory (DFT) approaches \cite{BS1,BS2,BS3,BS4,BS5,BS6,BS7,GammaFM1,GammaFM2,Herper99,BS8,Zhang11,BS9,BS10}, which allowed one to reproduce the experimental wave vector \cite{BS5,BS6,BS7} at the lattice parameter, corresponding to 
%$\gamma$-iron at 
low temperatures (or precipitates), while at sufficiently large lattice parameter the ferromagnetic phase was shown to be stable \cite{GammaFM1,GammaFM2,Herper99,BS8,Zhang11}. These first principle approaches allowed one also to obtain the lattice constant dependence of magnetic moment of $\gamma$-iron \cite{GammaFM1,GammaFM2,Herper99,Zhang11} and corresponding magnetic exchange interactions \cite{BS9,BS10}.

The {\it ab initio} DFT approaches do not allow, however, treating correlation, as well as temperature effects, which are often crucially important in strongly-correlated materials, such as iron. These effects may be especially pronounced in the presence 
%In particular, the formation 
of local magnetic moments, which 
%is often governed by 
appear in particular due to Hund's exchange interaction (in the so called Hund's metals \cite{Hund1,Hund2,Hund3,HundOur}).
%, can not be considered within the density functional theory alone. 
Recent dynamical mean-field theory (DMFT) studies \cite{OurGamma} have shown partly formed local moments in $\gamma$-iron at not very low 
%\com{(AB: 'at not very low' sounds a little unspecific) AK: It is explained by next sentences}
temperatures, allowing to consider it as a Hund's metal in some temperature range (see also Ref. \cite{HundOur}). In particular, the inverse local susceptibility is approximately linear in temperature above $T^*\sim 500$~K, corresponding to 
%. Therefore, this substance shows 
a crossover temperature scale from the local-moment to itinerant behavior.
%at the temperature scale $T^*$. 
%Interestingly, 
At temperatures below $T^*$ the local moments in $\gamma$-iron are screened by itinerant electrons. Indeed, fitting the inverse local susceptibility of Ref. \cite{OurGamma} at temperatures $T>T^*$ by the dependence $\chi_{\rm loc}^{-1}\propto T+2T_\textrm{K}$, determining the single-site Kondo temperature $T_\textrm{K}$, below which the local moments are screened \cite{Wilson}, yields $T_\textrm{K} \sim T^{*}$. At the same time, the local moments
%, being screened at low temperatures, 
%showing that in fact at temperatures below $T^{*}$ the local moments in this substance are screened by itinerant electrons, but likely 
do not strongly decay at not very low temperatures, which is confirmed by the calculated temperature dependence of dynamic local magnetic susceptibility \cite{OurGamma}.

On the other hand, due to thermal expansion, at high temperatures $\gamma$-iron is expected to exhibit stronger ferromagnetic, than antiferromagnetic correlations, as indicated by the DFT approaches \cite{GammaFM1,GammaFM2,Herper99,BS8,Zhang11} and experimental data \cite{GammaFechiq}.
%On the other hand, at sufficiently large lattice constant (and, therefore, in spite of the thermal expansion, sufficiently high temperatures) $\gamma$-iron is known to have more preferred ferromagnetic, than antiferromagnetic correlations \cite{GammaFM1,GammaFM2,GammaFechiq}. 
According to the comparison of the energies of antiferromagnetic and ferromagnetic phases in \textit{ab initio} studies (see, e.g., Refs. \cite{GammaFM1,GammaFM2,Herper99,BS8}), the transition between these phases occurs at the value of the lattice constant, corresponding to the temperature $T\sim 1000$~K, which is close to the %temperature 
% of
$\alpha$-$\gamma$ transition. 

Therefore, one can expect strong change of magnetic properties of $\gamma$-iron from itinerant antiferromagnet to local moment substance with dominating antiferromagnetic or ferromagnetic correlations with changing temperature.
%, and finally to the one with dominating ferromagnetic correlations. 
Although the dependence of the magnetic properties (in particular, exchange parameters) at zero temperature on lattice constant was studied previously within DFT calculations,
%~\cite{Herper99,Abrikosov07,Sjostedt02,Okatov11,Zhang11}, 
it seems important to investigate the effect of temperature and electronic correlations on magnetic properties of this substance. In the present paper we apply DFT+DMFT approach \cite{DFT+DMFT,Alpha_supercell} to study magnetic properties of $\gamma$-iron in a broad temperature range. In contrast to the previous study \cite{OurGamma} we vary lattice constant with changing temperature, and, more importantly, use the supercell DMFT approach, considered previously for $\alpha$-iron \cite{Alpha_supercell}, to extract momentum dependence of magnetic susceptibility and exchange interaction, including local vertex corrections. We find that indeed the character of magnetic fluctuations changes from dominating antiferromagnetic ones at low temperatures to ferromagnetic at the temperatures closer to the $\alpha$-$\gamma$ structural transition. We also obtain the corresponding magnetic exchange parameters.

The plan of the paper is the following. In Sect. II we discuss the method, in Sect. III present the results and finally in Sect IV we present conclusions.

\section{Method and computational details} \label{sec:computational_details}
%\vspace{-0.2cm}

\subsection{Supercell calculation of susceptibilities in DFT+DMFT }
\label{SectIIB}
%\vspace{-0.2cm}

%To obtain the non-uniform susceptibilities we use DFT+DMFT supercell approach.
%
First, 
we have performed DFT calculations using the full-potential linearized augmented-plane wave method implemented in the ELK code supplemented by the Wannier function projection procedure (Exciting-plus code).
%~\cite{elk}
%
The Perdew-Burke-Ernzerhof form of 
%generalized gradient approximation 
GGA was considered.
The calculations were carried out with the experimental temperature dependence of the lattice constant in the temperature range, where $\gamma$-iron exists in nature, ${a(T)=a_0 + a_1 T}$  where ${a_0=3.5519}$~\AA, $a_1=8.1593\! \times\! 10^{-5}$~\AA/K ~\cite{Seki2005}; in the following we extrapolate this dependence to lower and higher temperatures.
The convergence threshold for total energy was set to $10^{-6}$~Ry.
The integration in the reciprocal space was performed using 18$\times$18$\times$18\, $\textbf{k}$-point mesh for unit cell, while\, 15$\times$15$\times$15\,, and 12$\times$12$\times$12\, meshes were used for supercells with 2 and 4 atoms, respectively.
From converged DFT results we have constructed effective Hamiltonians in the basis of Wannier functions, which were built as a projection of the original Kohn-Sham states to site-centered localized functions as described in Ref.~\cite{Korotin08}, considering $3d$, $4s$ and $4p$ states.

In DMFT calculations we use the Hubbard parameter ${U\equiv F^0=4}$~eV
and Hund's rule coupling ${I\equiv (F^2+F^4)/14=0.9}$~eV,
%\com{(AB: $I \rightarrow J_S$?)}
where $F^0$, $F^2$, and $F^4$ are the Slater integrals as obtained in
Ref.~\cite{Belozerov_UJ} by the constrained DFT in the basis of $spd$ Wannier functions.
The on-site Coulomb interaction was considered in the density-density form.
The corresponding matrix of Hund's exchange can be expressed via the Coulomb interaction matrix $U^{mm'}_{\sigma,\sigma'}$ as $I^{mm'}=(U^{mm'}_{\sigma,-\sigma}%
-U^{mm'}_{\sigma,\sigma})(1-\delta_{mm'})$, $m$ and $\sigma$ 
are orbital 
and spin 
indexes. 
The double-counting correction was taken in the fully localized limit.
The impurity problem was solved by the hybridization expansion continuous-time quantum Monte Carlo method~\cite{CT-QMC}. 
In our calculations we neglect the redistribution of charge density on the DFT level caused by the self-energy from DMFT, since iron is a moderately correlated metal, in which the $3d$ states are only weakly hybridized with $4s$ and $4p$ states; previous charge self-consistent studies of iron 
%performed 
%within the charge self-consistent implementation 
(e.g., Refs.~\cite{Pourovskii2013,Kvashnin2015})
%, which however 
did not result in any significant discrepancies with other DFT+DMFT studies.

The non-uniform static spin susceptibility 
%\vspace{-0.0cm}
\begin{equation}
\chi^{mm'}_{\mathbf q}=\frac{1}{N}\int_0^\beta d\tau \sum_{ij} \langle s^z_{im}(0) s^z_{jm'}(\tau)  \rangle e^{i {\bf q}({\bf R}_j-{\bf R}_i)},
%\vspace{-0.2cm}
\end{equation}
where ${\mathbf s}_{im}=c^{+}_{im\sigma} \mbox{\boldmath $\sigma $}_{\sigma \sigma'} c_{i m \sigma'}/2$ 
are electronic spin operators, $c_{im\sigma} (c^{+}_{im\sigma}$) are electron destruction (creation) operators
%, corresponding to the electronic degrees of freedom 
($i$ is the 
%, $m$ and $\sigma$ are
site
%-, orbital 
%and spin 
index), 
%\mbox{\boldmath $\sigma $} are the Pauli matrices, 
can be obtained by calculating a response to a small staggered external field introduced in the DMFT part in a suitable supercell. 
%when calculating the local Green's function
Namely, for the orbital-resolved magnetic susceptibility we have $\overline{\chi}_{\mathbf{Q}_i}^{mm^{\prime}}
=4 \mu_B^2 {\chi}_{{\mathbf{Q}_i}}^{mm^{\prime}}=
\partial M_{\mathbf{Q}_i}^{m^{\prime}}/\partial H_{\mathbf{Q}_i}^m$, where 
%$M_{\mathbf{q}_{i}}^{m^{\prime}}$ is the magnetization of orbital $m^{\prime}$,
$H_{\mathbf{Q}_i}^m$ is the magnetic field applied to orbital~$m$ and
corresponding to the wave vector $\mathbf{Q}_i$,
$M_{\mathbf{Q}_i}^{m^{\prime}}$ is the magnetization of orbital $m^{\prime}$.
In the real space, the applied field
%can be expressed as
takes a form
${\mathbf{H}_{\mathbf{R}_j}^{m,i} = \mathbf{H}_0\, \textrm{cos}(\mathbf{Q}_i \mathbf{R}_j)}$,
where 
$\mathbf{R}_j$ is the position vector of site $j$,
$\mathbf{H}_0$ is a constant small field.
%value chosen to provide a linear response. 
In practice, we have used the magnetic field corresponding to splitting of the single-electron energies by 0.02 eV.  This field was checked to provide a linear response and was considered to be small enough to neglect the redistribution of charge density on the DFT level. %We note that our DFT+DMFT calculations were performed without full charge self-consistency.

For high-symmetry wave vectors the corresponding supercells are compact,
and therefore can be studied by 
%, which makes them appropriate for treatment by
%DMFT with several impurity problems. 
%multi-impurity extension
real-space extension
%\com{(AK: Is the name correct? Usually multi-impurity problem means problems with coupled impurities. AB: It is better to replace by "inhomogeneous (or real-space) DMFT". At least, this name is used in several recent papers.)}
of DMFT (see, e.g., Refs. ~\cite{Potthoff1999_1,Potthoff1999_2}).
In this extension, the self-energy is still local but assumed to be site dependent.
As a result, several single-impurity problems have to be solved at each self-consistency loop. Note that neglect of the non-local components of the self-energies may yield an underestimate of the non-local components of the susceptibility. We expect, however, that because of strong on-site electronic correlations, non-local components of the self-energy do not change substantially the obtained results.

%This approach has been successfully applied to a Hubbard-type model with disorder~\cite{} and inhomogeneous systems~\cite{Potthoff1999}.

To calculate the non-uniform susceptibilities we have constructed supercells containing up to four atoms and corresponding to seven high-symmetry points.
In particular, for wave vector ${\mathbf{Q}_{\textrm{X}_1}=(0,0,2\pi)/a}$
we considered supercells containing two nearest-neighbor atoms at $(0,0,0)$ and $(0,a/2,a/2)$ in Cartesian coordinates with lattice vectors 
${\{0,a,0\}}$,
${\{0,0,a\}}$, and 
${\{a/2,a/2,0\}}$.
The same atoms were used to construct a supercell for ${\mathbf{Q}_{\textrm{L}}=(\pi,\pi,\pi)/a}$ with lattice vectors 
${\{a,a,0\}}$,
${\{a/2,-a/2,0\}}$, and
${\{0,-a/2,a/2\}}$.
For ${\mathbf{Q}_{\textrm{W}_1}=(\pi,2\pi,0)/a}$, we built a supercell with four atoms by including two extra atoms at $(a,0,0)$ and $(-a/2,a/2,0)$.
The lattice vectors for this supercell are 
${\{a,a/2,a/2\}}$,
${\{2a,0,0\}}$, and 
${\{0,a,0\}}$.
For ${\mathbf{Q}_{\textrm{X}_2}=(2\pi,0,0)/a}$, ${\mathbf{Q}_{\textrm{X}_3}=(0,2\pi,0)/a}$, ${\mathbf{Q}_{\textrm{W}_2}=(2\pi,\pi,0)/a}$, and ${\mathbf{Q}_{\textrm{W}_3}=(2\pi,0,\pi)/a}$ the supercells have been constructed in a similar manner by permutation of corresponding components.
The orbital-resolved results for the wavevectors  ${\bf Q}_{\textrm{X}_i}$ and ${\bf Q}_{\textrm{W}_i}$ with different $i$ are not equivalent because of the orientation of $d$-orbitals in certain directions in real space. Their rotation by point group operations than yields the off-diagonal (in the orbital space) components of the spin operators ${\mathbf s}_i^{mm'}=c^{+}_{im\sigma} \mbox{\boldmath $\sigma$}_{\sigma \sigma'} c_{i m' \sigma'}$/2, yielding non-Heisenberg components of the exchange interaction, which are not considered here. 
%\newpage

%\vspace{-0.3cm}
\subsection{Formulas for magnetic exchange}
%\vspace{-0.3cm}

The %corresponding 
orbital-resolved exchange interaction $\mJ_{ij}^{mm'}$ can be represented in the RKKY-like form, its Fourier transform reads \cite{Alpha_supercell}%
\begin{equation}
{\mathcal J}_{\mathbf{q}}^{mm^{\prime}}=2I^{mm^{\prime\prime}}\left(
\chi_{\mathbf{q}}^{m^{\prime\prime}m^{\prime\prime\prime}}\right)
_{\mathrm{irr}}I^{m^{\prime\prime\prime}m^{\prime}},%
\label{Eq:Jq}
\end{equation}
where the summation (i.e. matrix product) over repeated indexes is assumed and 
%For the following, we introduce 
the (transverse) irreducible parts of non-uniform electronic susceptibilities $(\chi^{mm'}_{\mathbf q})_{\rm irr}$ are related to the magnetic susceptibilities $\chi^{mm'}_{\mathbf q}$
%\begin{equation}
%(\chi^{mm'}_{\mathbf q})=\frac{1}{N}\int_0^\beta d\tau \sum_{ij} \langle s^z_{im}(0) s^z_{jm'}(\tau)  \rangle e^{i {\bf q}({\bf R}_j-{\bf R}_i)},
%\end{equation}
%where ${\mathbf s}_{im}=c^{+}_{im\sigma} \sigma_{\sigma \sigma'} c_{i m \sigma'}$ are electronic spin operators, 
by the Hund's exchange interaction, 
%matrix $I^{mm^{\prime}}$ 
%are introduced:%
\begin{equation}
\left(  \chi_{\mathbf{q}}^{mm^{\prime}}\right)  _{\mathrm{irr}}=\left[
\left(  2\chi_{\mathbf{q}}^{mm^{\prime}}\right)  ^{-1}+I^{mm^{\prime}%
}\right]  ^{-1},%
\label{chi_irr}
\end{equation}
$\left[  ...\right]  ^{-1}$ denotes the matrix inverse with respect to 
the orbital indexes, the factor of $2$ accounts the difference between the
transverse and longitudinal susceptibilities.
%
%

%To calculate the resulting (physical) exchange interaction, non-local
%corrections to the Curie temperature and the energy, we consider 
%The effective
%model for the spin degrees of freedom ${\bf S}$ can be obtained from the action 
%\begin{equation}
%S=S_{\rm el}[c,c^{+}]+\int d\tau H_H[S,s],
%\end{equation}
%where $H_H[S,s]$ is given by the equation (\ref{HH}) and $S_{\rm el}[c,c^{+}]$ includes the kinetic term and all interelectron interactions, apart from the Hund one. 

While the components of the exchange interaction ${\mJ}_{\mathbf{Q}_{i}}$ can be determined from the obtained irreducible susceptibilities, to
interpolate between different points $\mathbf{Q}_{i}$  we consider an expansion 
\begin{align}
{\mathcal J}_{\mathbf{q}}^{mm^{\prime}} &  ={\mathcal J}^{mm^{\prime},(0)}+\overline{\mathcal J}_{\mathbf{q}%
}^{mm^{\prime}},\label{EqJ1}\\
\overline{\mathcal J}_{\mathbf{q}}^{mm^{\prime}} &  ={\mathcal J}^{mm^{\prime},(1)}_{xy}\cos(aq_{x}/2)\cos
(aq_{y}/2)\notag \\
&+\mJ^{mm^{\prime},(1)}_{xz}\cos(aq_{x}/2)\cos
(aq_{z}/2)\notag\\
&+\mJ^{mm^{\prime},(1)}_{yz}\cos(aq_{y}/2)\cos
(aq_{z}/2)\nonumber\\
&  +\mJ^{mm^{\prime},(2)}_x \cos(aq_{x})+\mJ^{mm^{\prime},(2)}_y\cos(aq_{y})\notag\\
&+\mJ^{m m^{\prime},(2)}_z\cos(a q_{z})+\mJ^{mm^{\prime},(3)}\left[\cos(aq_{x})\cos(aq_{y})\right.\notag\\
&+\left.\cos(aq_{y})\cos(aq_{z})
+\cos(aq_{z})\cos(aq_{x})\right],\label{EqJ2}
\end{align}
such that $\mJ^{mm^{\prime},(r)}$ are determined by $\mJ_{\mathbf{Q}_{i}%
}^{mm^{\prime}}.$ To determine eight matrices $\mJ^{mm',(0,3)}$, $\mJ_{ab}^{mm',(1)}$, and $\mJ_{a}^{mm',(2)}$ we consider irreducible susceptibilities for eight wave vectors ${\bf Q}_\Gamma=(0,0,0)$, ${\bf Q}_L$,
%=(\pi,\pi,\pi)$, 
${\bf Q}_{X_i}$, and ${\bf Q}_{W_i}$
%=(0,0,2\pi)$, ${\bf Q}_{X_2}=(2\pi,0,0)$, ${\bf Q}_{X_3}=(0,2\pi,0)$, ${\bf Q}_{W_1}=(\pi,2\pi,0)$, ${\bf Q}_{W_2}=(2\pi,\pi,0)$, ${\bf Q}_{W_3}=(2\pi,0,\pi)$. The orbital-resolved results for the wavevectors  ${\bf Q}_{X_i}$ and ${\bf Q}_{W_i}$ with different $i$ are not equivalent because of the orientation of $d$-orbitals in certain directions in real space. In principle, their rotation than yields the off-diagonal (in the orbital space) components of the spin operators ${\mathbf s}_i^{mm'}=c^{+}_{im\sigma} \sigma_{\sigma \sigma'} c_{i m' \sigma'}$, which are not considered here. 
Because of neglect of the off-diagonal spin operators ${\mathbf s}_i^{mm'}$, the present treatment is only approximate; as we will see in the following, however, the crystal symmetry breaking in final results for the exchange interaction is sufficiently small, and can be neglected. 
	To extract the physical exchange from the obtained matrices $\mJ^{mm',(i)}_{ab}$, we calculate it as 
\begin{equation} 
 J^{(i)}_{ab}=\sum_{mm'} \mJ^{mm',(i)}_{ab} \mu^2_{mm'}/\sum_{mm'} \mu^2_{mm'},
 \label{JiAv}
 \end{equation} 
 where $\mu^2_{mm'}=3[d(1/\chi^{mm'}_{\rm loc})/dT]_a^{-1}$ is the matrix of squares of local moments, $\chi^{mm'}$ are orbital-dependent local susceptibilities, index $a$ indicates that the lattice constant is kept constant when evaluating the derivative.

\section{Results and discussion}

\begin{figure}[t]
\includegraphics[width=0.454\textwidth]{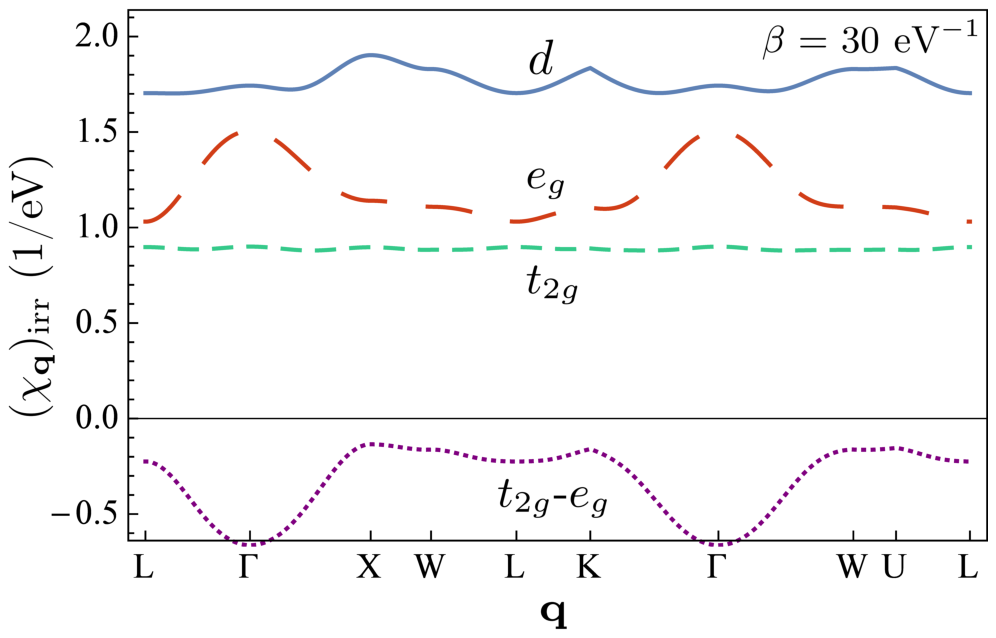}\vspace{0.3cm}
\includegraphics[width=0.46\textwidth]{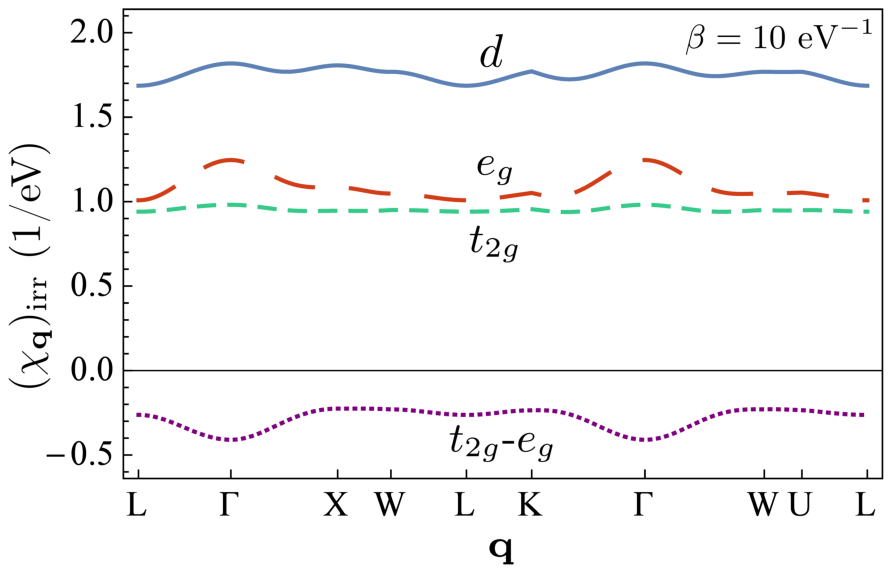}
\caption{(Color online)
\label{Fig:chi_irr}
%Top: 
Momentum dependence of the irreducible susceptibility, summed over all orbitals (blue solid line), $t_{2g}$ orbitals (green short-dashed line), $e_g$ orbitals (red long-dashed line), and $t_{2g}$-$e_g$ contributions (purple dotted line) for $\beta=30$ eV$^{-1}$ ($a=3.583$~\AA, top) and $\beta=10$ eV$^{-1}$ ($a=3.647$~\AA, bottom).
%\com{(AB: We could add labels "t2g", "eg", etc to the figures.) AK: Ok, I'll do that. Yes, but then I need to send you a data, I have them in analytic, and not numeric form in Mathematica. I can add directly to eps files. Ok, also eV should be added to vertical axes, and q made not italic. AB: Ok.}
}
\end{figure}

In Fig.~\ref{Fig:chi_irr} we present the resulting momentum dependencies of the irreducible susceptibilities, summed over all and part of the orbitals; the interpolation between symmetric points is performed by calculating the exchange interactions in Eqs.~(\ref{EqJ1}) and (\ref{EqJ2}) and inverting then Eq. (\ref{Eq:Jq}). Although the obtained dependences are qualitatively similar to those obtained earlier from the bare bubble in DMFT \cite{OurGamma}, the numerical values of susceptibilities are approximately two times larger (similarly to previous study of $\alpha$-iron \cite{Alpha_supercell}) because of the vertex corrections. At low temperatures ($\beta=30$ eV$^{-1}$) the maximum of the obtained susceptibility is at the X point, which shows dominant antiferromagnetic correlations. The susceptibility is, however, weakly momentum dependent, such that these correlations compete with fluctuations with other wave vectors, in particular $\Gamma$ (i.e. ferromagnetism), W, and K. As can be seen from partial contributions, the weak momentum dependence appears as a result of compensation of $e_g$ and mixed $t_{2g}$-$e_g$ contributions, while the $t_{2g}$ contribution is almost momentum-independent itself. For $\beta=10$ eV$^{-1}$ the momentum dependence of total irreducible susceptibility becomes even weaker; the susceptibility at $\Gamma$ point becomes close to that at the X point, which shows that ferromagnetic correlations are as strong, as the antiferromagnetic ones at this temperature. 
%slightly dominate at this temperature. 
Note that weak momentum dependence of the magnetic susceptibility and close competition of ferro- and antiferromagnetic correlations at temperatures $T\sim 1200$~K, at which $\gamma$-iron exists in nature, agrees with the experimental results \cite{GammaFechiq}. 
%This qualitative change of the 
Such a momentum dependence of the susceptibility at not too low temperatures, which is qualitatively different from the low-temperature behavior, 
%with temperature is 
is obtained entirely due to using supercell DMFT approach, accounting for the vertex corrections for the magnetic susceptibility, and it was not found in the calculation of momentum dependence of bubble of Green functions in DMFT approach of Ref. \cite{OurGamma}. 

\begin{figure}[t]
%\includegraphics[width=0.456\textwidth]{J_beta30.eps}\vspace{0.3cm}
%\includegraphics
%[width=0.456\textwidth]
%{FigchiT.eps}%\vspace{0.3cm}
%{uniform_and_staggered_susc.eps}%\vspace{0.3cm}
%\\
%\includegraphics
%[width=0.456\textwidth]
%{FigchiT.eps}%\vspace{0.3cm}
%{local_sus_.eps}%\vspace{0.3cm}
%\includegraphics[width=0.456\textwidth]{Fig2_susc_with_local.eps}
\includegraphics[width=0.456\textwidth]{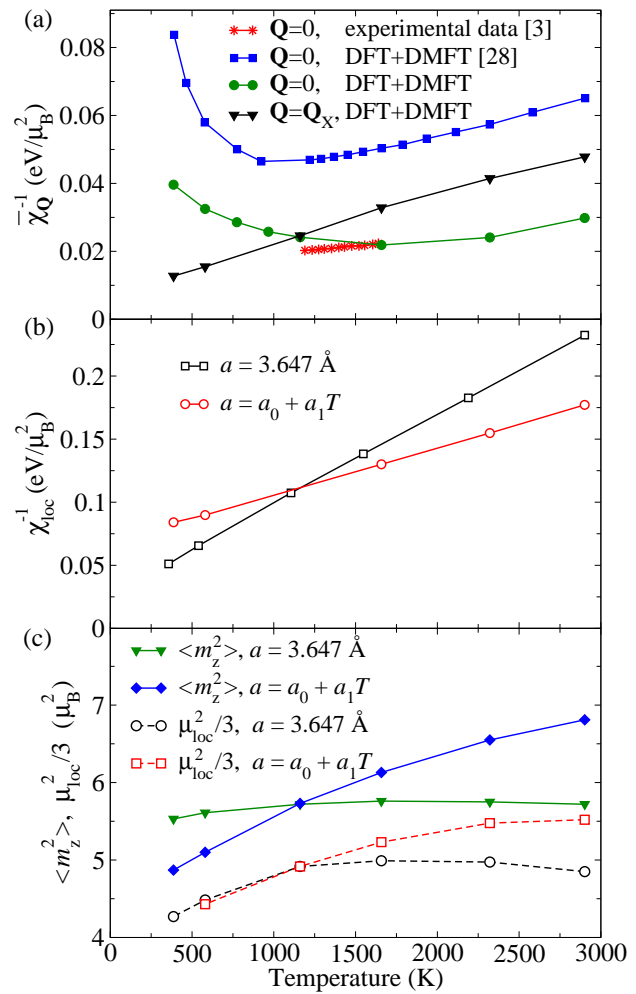}
\caption{(Color online)
\label{Fig:chiQT}
%Top: 
Temperature dependencies of the inverse magnetic uniform and staggered susceptibilities (top panel), obtained within the (supercell) DFT+DMFT approach together with the experimental data for the uniform susceptibility. The inverse local magnetic susceptibility is shown in the middle panel. The instantaneous average ${\langle m_z^2\rangle}$ and local magnetic moments from local susceptibility are shown in the bottom panel.}
%{\bf AK: Please restrict the old data by 3000K, and replace references by Ref. [...] (the latter have to be verified/inserted in the end)}
%; (b) and (c) show inverse irreducible uniform and staggered susceptibilities $\sum_{m,m'} (\chi_{\mathbf Q}^{m,m'})_{\rm irr}$, respectively; dashed lines show contributions of $m,m'$ belonging to $t_{2g}$ and $e_g$ orbitals. 
%{\bf AK: What about joining two plots to have common temperature axis at the bottom (i.e. remove temperature captions from the upper plot?) Also please insert a) b). Should be the height of second plot made smaller?}}
\end{figure}

%\begin{figure}[b]
%\includegraphics
%[width=0.4\textwidth]
%{local_sus_.eps}%\vspace{0.3cm}
%\caption{(Color online)
%\label{Fig:chi_loc}
%Temperature dependences of the inverse local magnetic susceptibility, obtained within the DFT+DMFT approach.
%}
%\end{figure}

The temperature dependence of the uniform and staggered susceptibilities $\overline{\chi}_{\mathbf Q}=\sum_{m,m'} \overline{\chi}_{\mathbf Q}^{m m'}$, corresponding to ${\mathbf Q}=0$ and  ${\mathbf Q}={\mathbf Q}_X$, respectively, is shown in Fig. \ref{Fig:chiQT}(a) (as mentioned above, the susceptibilities, corresponding to different ${\mathbf Q}={\mathbf Q}_{X_i}$ are slightly different; the difference is however small). In agreement with previous calculations \cite{OurGamma} the inverse uniform susceptibility decreases with increasing temperature at low temperatures $T$. The value of the inverse uniform susceptibility, obtained in the present study, is approximately twice smaller than found previously \cite{OurGamma}, mainly due to larger (and more realistic) choice of the Coulomb interaction  and agrees well with the experimental data. The slope of the temperature dependence of the inverse uniform susceptibility near the experimental temperature of $\alpha$-$\gamma$ structural transition is not obtained correctly, but one should take into account that the Curie and structural transition temperatures are overestimated in the considered theory, treating Ising symmetry of Hund's exchange \cite{Leonov}. The obtained slope of the inverse susceptibility at the expected theoretical temperature of $\alpha$-$\gamma$ transition $1.2T_{\textrm{C},\alpha}^{\rm LDA+DMFT}\simeq 2600$~K (the Curie temperature of $\alpha$-iron $T_{\textrm{C},\alpha}^{\rm LDA+DMFT}$, obtained within LDA+DMFT analysis, was taken from Refs. \cite{Sangiovanni,Alpha_supercell}), yields better agreement with the experimental data for the slope.

On the other hand, the staggered susceptibility increases with decreasing temperature, and approximately fulfills the Curie-Weiss law. The corresponding Weiss temperature $\theta_{\rm stagg}\approx -340$~K is, however, negative, such that no long-range magnetic order is obtained at low temperatures (at least from the extrapolation of the obtained inverse susceptibility).
%above \com{(AB: below? AK: from the temperatures above. AB: Is the phrase 'above room temperature' necessary? To me it seems ambiguous.)} room temperature). 
The long-range order in copper precipitates may occur due to temperature dependence of the lattice constant, somewhat different from the considered one, the surface/volume anisotropy effects of $\gamma$-iron nanoparticles, %\cite{FeAnis}, 
as well as the anisotropic dipole-dipole interaction.
It is important to note that despite the negative Weiss temperature, both ferro- and antiferromagnetic correlations at $T\sim 1000$~K are sufficiently strong. In particular, the corresponding values of the inverse staggered and uniform susceptibilities are comparable to the uniform susceptibility of $\alpha$-phase at the $\alpha$-$\gamma$ transition temperature, as follows from 
%the continuity of the susceptibility at the structural transition \com{(AB: Experimentally, there is a significant jump of uniform susceptibility at the $\alpha$-$\gamma$ transition~\cite{Susc3})} and 
previous theoretical results of uniform susceptibility of $\alpha$-iron \cite{Alpha_supercell} at   $T=1.2T_{\textrm{C},\alpha}^{\rm LDA+DMFT}$.
%,  obtained previously in Ref. . 
%Therefore, the dynamic mean-field theory seem to underestimate the tendency to antiferromagnetic order (why is it so? usually it overestimates??).

%The uniform and staggered irreducible susceptibilities (see Figs. \ref{Fig:chiQT}b,c) show qualitatively the same temperature dependences, as the total susceptibilities of Fig. \ref{Fig:chiQT}a with weaker temperature dependence. The orbital-resolved contributions show Curie-Weiss like behavior of $e_g$ contributions, while the $t_{2g}$ states yield irreducible susceptibilities, which decrease with decreasing temperature. 

The temperature dependence of the inverse local susceptibility is shown in Fig.~\ref{Fig:chiQT}(b). In agreement with previous results \cite{OurGamma} in the considered temperature range it 
is approximately linear 
%temperature dependence 
for both, fixed and temperature-dependent lattice constant.
%up to the 
%starts to deviate from the linear dependence at the 
%temperature $T^{*}\sim 500$ K. 
The temperature dependencies of the instantaneous and static local moments, extracted from the average $\langle m_z^2\rangle$ (which is almost site-independent due to site-diagonal form of the self-energy), where $m_z=2 \mu_B \sum_m s^z_m$, and the derivative of the inverse susceptibility  $\mu^2_{\rm loc}=3[d(1/\chi_{\rm loc})/dT]_a^{-1}$ obtained from
%is the total spin projection 
 $\chi_{\rm loc}=\sum_{m,m'}\chi^{mm'}_{\rm loc}$, respectively, 
%is the total local susceptibility, 
are shown in Fig.~\ref{Fig:chiQT}(c). One can see that at fixed lattice constant the average $\langle m_z^2\rangle$ is weakly temperature dependent; the local moment  $\mu^2_{\rm loc}$ shows somewhat stronger temperature dependence, especially at low temperatures, 
%$T<1000$~K, 
reflecting a tendency of destroying static local moments at lower temperatures \cite{OurGamma}. The suppression of the local moments is not pronounced in the considered temperature range, and, therefore,
%In the considered temperature range the deviation is, however, small, so that 
they are well formed above the lowest considered temperature $T=1/30$ eV.  The same characteristics of local moments, calculated with temperature dependent lattice constant, show stronger temperature dependencies, reflecting effect of changing lattice constant. At not too high temperatures $T<1500$~K we find weak effect of the lattice constant change on $\mu_{\rm loc}^2$. The obtained value of magnetic moment $\mu_{\rm loc}\approx 3.8\mu_B$ at $T=1200$~K agrees with previous DMFT study \cite{OurGamma}, but somewhat larger than that obtained in DFT approach in both, low-spin (antiferromagnetic) and high-spin (ferromagnetic) phases \cite{GammaFM1,GammaFM2,Herper99}. %Comparing however with DFT
On the other hand, for the saturated magnetic moment $\mu_{\rm sat}$, defined by $\mu_{\rm loc}^2=\mu_{\rm sat}(\mu_{\rm sat}+2\mu_B)$, we find the value $\mu_{\rm sat}\approx 2.9\mu_B$, which is closer to the high-spin state DFT result. %and paramagnetic \cite{Zhang11} phases.

\begin{figure}[b]
\vspace{0.3cm}
\vspace{-0.5cm}
\includegraphics[trim=0cm 0cm 0cm 0cm,width=0.47\textwidth]{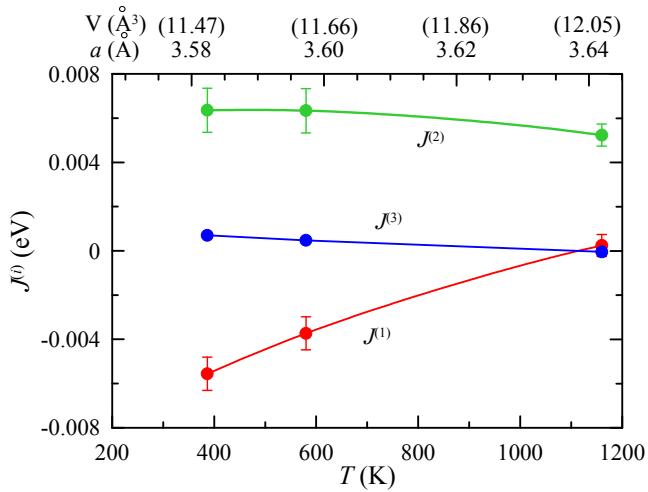}%\vspace{0.3cm}
\caption{(Color online)
\label{Fig:JT}
Temperature dependence of the magnetic exchange integrals $J^{(i)}$ in first three coordination spheres; the upper axis shows respective lattice constants and unit cell volumes for the considered temperatures. The error bars show only uncertainty, related to the Heisenberg form of magnetic interaction, see text.   
}
\end{figure}

\begin{figure}[t]
\includegraphics[width=0.456\textwidth]{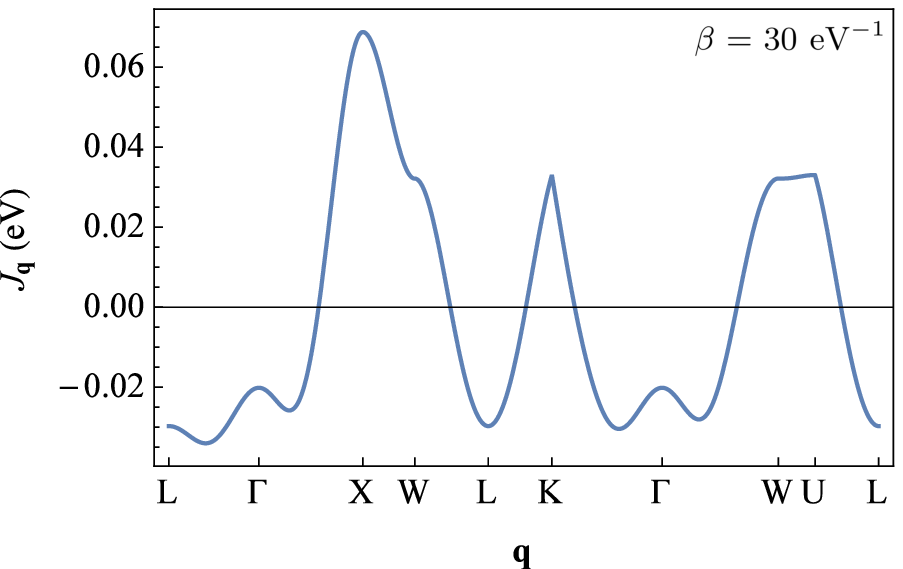}\vspace{0.3cm}
\includegraphics[width=0.456\textwidth]{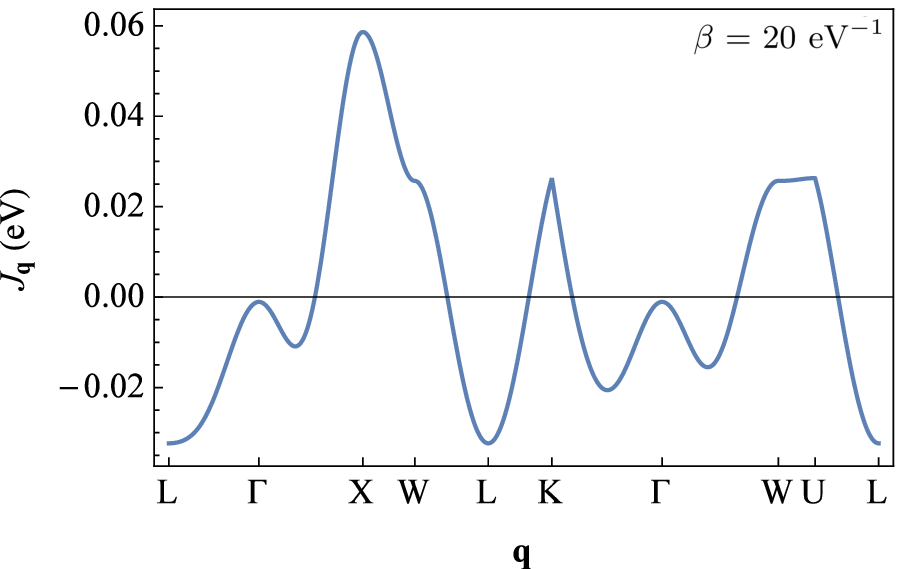}\vspace{0.3cm}
\includegraphics[width=0.456\textwidth]{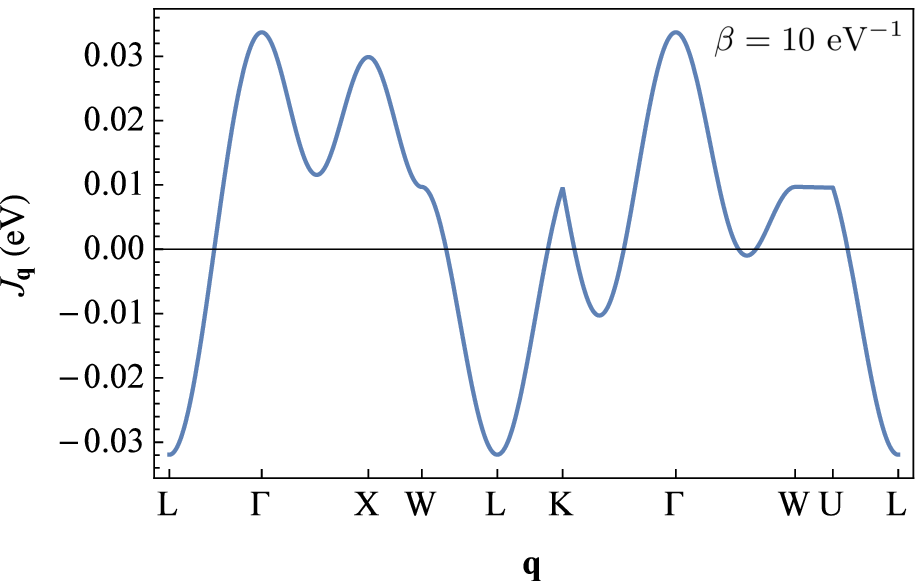}
\caption{(Color online)
\label{Fig:Jq}
%Top: 
Momentum dependence of the magnetic exchange integral for $\beta=30$ eV$^{-1}$ (top), $\beta=20$ eV$^{-1}$ (middle), and $\beta=10$ eV$^{-1}$ (bottom).
%{\bf AK: Please make $J_q$ with lower q index, add eV, and q not italic}
}
\end{figure}

Let us consider the results for the magnetic exchange. The temperature dependence of the orbital-averaged exchange parameters $J^{(i)}$, obtained according to the Eq.~(\ref{JiAv}), is shown in Fig.~\ref{Fig:JT} (we average the results also with respect to the space indexes $a,b$ and show the corresponding spread of the obtained values for different $a,b$ by error bars, which correspond physically to assuming Heisenberg form of the exchange interaction as discussed above; the on-site contribution $J^{(0)}\approx 0.96$~eV is obtained rather weakly temperature dependent). One can see that the exchange $J^{(3)}$ remains small in the considered temperature range and  the dominant contribution comes from first two coordination spheres. In this situation (provided $J^{(2)}>0$) the type of the ground state magnetic configuration (and dominant magnetic correlations at finite temperature) is determined by the sign of $J^{(1)}$: it is ferromagnetic for  $J^{(1)}>0$ and antiferromagnetic with the wave vector $(0,0,2\pi)/a$ for $J^{(1)}<0$. One can see that the nearest-neighbor exchange is antiferromagnetic at low temperatures and favors the $(0,0,2\pi)/a$ short-range order, in agreement with the analysis of susceptibilities (weak deviations from the wave vector ${\bf Q}_X$ can not be treated in the considering supercell approach). Approaching the temperature $\beta=10$~eV$^{-1}$, which is closer to the $\alpha$-$\gamma$ structural transition, we obtain, however, almost vanishing nearest neighbor exchange, such that the system appears on the boundary between the regimes with strong ferro- and antiferromagnetic correlations, also in agreement with the analysis of the susceptibility above. We note that DFT calculations yield change of sign of nearest-neighbor exchange at the unit cell volumes 11.8~\AA$^3$ \cite{BS10} or 11.4~\AA$^3$ \cite{BS9}, which are substantially smaller than the unit cell volume $V_0=12.1$~\AA$^3$ at $\beta=10$~eV$^{-1}$.
%, although the corresponding estimate from the total energy yields somewhat larger values \cite{GammaFM1,Herper99,BS8}. In the considered theory, which treats effective Heisenberg exchange between magnetic moments the transitions from the exchange integrals and free energy 
Therefore, present theory allows one to obtain better agreement with the experimental data of Ref. \cite{GammaFechiq}.
%and the experimental data of Ref. \cite{GammaFechiq}. 
%Obtained small value of third nearest neighbor magnetic exchange does not change the temperature of the crossover between ferro- and antiferromagnetic correlations; a
Although longer-range than third neighbors magnetic exchanges are not considered in the present approach (and third neighbor exchange is small), also in DFT calculations \cite{BS9,BS10} the third- and longer range magnetic exchanges almost compensate each other in the vicinity of the unit cell volume $V_0$. 
%12.1~\AA$^3$, corresponding to the temperature $\beta=10$~eV$^{-1}$. 
%At the same time, 

The resulting momentum dependence of the magnetic exchange $J_{\bf q}$, calculated analogously to $\overline{\mJ}_{\mathbf{q}}$ in Eq.~(\ref{EqJ2}) with the obtained exchange integrals $J^{(i)}$, substituted instead of $\mJ^{(i)}_{mm'}$, is shown in Fig. \ref{Fig:Jq}. In agreement with the above discussed results, we obtain $J_{{\bf Q}_X}>J_0$ at low temperatures and $J_{{\bf Q}_X}\approx J_0$ at $\beta=10$ eV$^{-1}$. The magnetic exchange $J_0=0.032$ eV at $\beta=10$ eV$^{-1}$ (which in our approach is provided mainly by the next-nearest-neighbor interaction), multiplied by the square of effective spin $3/2$ (corresponding to our magnetic moment $\mu_{\rm loc}^2\approx 15\mu_B^2$) is comparable (but somewhat larger) than the exchange $J_0=0.05$ eV between unit spin vectors, obtained in recent DFT approach  \cite{BS9}.

\section{Conclusion} \label{sec:conclusions}

We have studied magnetic properties and magnetic exchange interactions in paramagnetic fcc iron
by a combination of density functional theory and dynamical mean-field theory (DFT+DMFT).
By using supercell approach and interpolating the values of magnetic susceptibility between the symmetric points of the Brillouin zone with the  expansion of the magnetic exchange in coordination spheres up to third nearest neighbors, we have obtained weak momentum dependence of the magnetic susceptibility. In agreement with previous theoretical results and the experimental data we find that the antiferromagnetic correlations with the wave vector close to $(0,0,2\pi)/a$ dominate at low temperatures. At the same time, antiferromagnetic and ferromagnetic correlations closely compete at the temperatures $T\sim 1000$~K, where $\gamma$-iron exists in nature. Although this latter result is also in agreement with the experimental data \cite{GammaFechiq}, to our knowledge it has not been reproduced theoretically previously. The analysis of the inverse uniform susceptibility shows improvement of the agreement with the experimental data in comparison with previous theoretical study due to more realistic Coulomb interaction; the obtained  
inverse staggered susceptibility shows linear temperature dependence at low temperatures, with negative Weiss temperature $\theta_{\rm stagg} \approx -340$~K. The inverse local susceptibility is found to be also linear at not too low temperatures, showing well formed local moments. Analysis of magnetic exchange between these local moments shows that the dominant contribution to the magnetic exchange comes from first two coordination spheres; the nearest-neighbor exchange is found to be antiferromagnetic at low temperatures, while at temperature of the $\alpha$-$\gamma$ structural phase transition its absolute value becomes small, and the system appears on the boundary between the regimes with strongest antiferro- and ferromagnetic correlations. At higher temperatures the nearest- and next-nearest exchanges are ferromagnetic. We note that in our study the crossover between the regimes with strongest ferro- and antiferromagnetic correlations is due to change of preferred orientation of local moments with weakly varying size $\mu_{\rm loc}$, which is in contrast to  the transition from low- to high spin itinerant state in DFT. In our calculations we have used the density-density form of Hund's exchange, which was shown to significantly overestimate the $\alpha$-$\gamma$ structural phase transition temperature \cite{Leonov,our_alpha_gamma}. However, our results are expected to remain qualitatively unchanged for the SU(2) symmetric form, since at high temperatures the ferromagnetic correlations are found to be strongly pronounced.
%Although we have used the density-density form of Hund's exchange in the DMFT calculations of the paper, the results are expected to remain qualitatively unchanged for the SU(2) symmetric form. 
%In particular, we expect the crossover temperature from the regime of strongest antiferro- to ferromagnetic correlations only weakly change for SU(2) form of Hund's exchange.

The obtained results extend and deepen previous understanding of the magnetic properties of $\gamma$-iron and stress important role of ferromagnetic correlations in this substance at not too low temperatures. Although the ferromagnetic instability at large lattice parameter was studied previously within band structure calculations \cite{GammaFM1,GammaFM2,Herper99,BS8,Zhang11}, using dynamical mean-field theory allows us to consider the evolution of magnetic properties with raising temperature and describe their change from $\gamma$-iron in Cu precipitates at low temperatures to the $\gamma$-iron, existing in nature. The obtained close competition of ferro- and antiferromagnetic correlations (including possible phase separation on the short-range ordered ferro- and antiferromagnetic regions) may also help to explain the anti-Invar behavior of $\gamma$-iron, beyond high- and low-spin states mechanism, proposed previously \cite{AIGamma}. 

\newpage
\begin{acknowledgments}
The work was supported by the Russian Science Foundation (Project No. 14-22-00004).
\end{acknowledgments}

% ======================================================

\end{document}